\documentclass{aa}  

\usepackage{graphicx}
\usepackage{epsfig}
\usepackage{txfonts}
\begin{document}

   \title{Difference image analysis: The interplay between the photometric scale factor and systematic 
          photometric errors}

%   \subtitle{}

   \author{D.M. Bramich\inst{1}
          \and
          E. Bachelet\inst{1}
          \and
          K.A. Alsubai\inst{1}
          \and
          D. Mislis\inst{1}
          \and
          N. Parley\inst{1}
%          C. Ptolemy\inst{2}\fnmsep\thanks{Just to show the usage
%          of the elements in the author field}
          }

   \institute{Qatar Environment and Energy Research Institute, Qatar Foundation, P.O. Box 5825, Doha, Qatar\\
              \email{}
%         \and
%             University of Alexandria, Department of Geography, ...\\
%             \email{c.ptolemy@hipparch.uheaven.space}
%             \thanks{The university of heaven temporarily does not
%                     accept e-mails}
             }

   \date{Received March 15, 2015; accepted March 16, 2015}

% \abstract{}{}{}{}{} 
% 5 {} token are mandatory
 
  \abstract
   {Understanding the source of systematic errors in photometry is essential for their calibration.}
   {We investigate how photometry performed on difference images can be influenced by errors in the photometric scale factor.}
   {We explore the equations for difference image analysis (DIA) and we derive an expression describing how errors in the difference flux, the photometric scale factor
    and the reference flux are propagated to the object photometry.}
   {We find that the error in the photometric scale factor is important, and while a few studies have shown that it can be at a significant level, it is
    currently neglected by the vast majority of photometric surveys employing DIA.}
   {Minimising the error in the photometric scale factor, or compensating for it in a post-calibration model, is crucial for reducing the systematic
    errors in DIA photometry.}


   \maketitle

\section{Introduction}

The technique of difference image analysis (DIA; \citealt{ala1998}; \citealt{ala2000}; \citealt{bra2008}; \citealt{bec2012}; \citealt{bra2013})
is based on matching a reference image to a target image by modelling the differences in alignment, point-spread function (PSF),
exposure time, atmospheric extinction, and sky background between them. Specifically, a convolution kernel is used
to model the first four differences (to within a small translational offset) while an additive differential background is used to model
the last. The reference image is transformed by convolving it with the kernel and adding the differential background,
and the result is subtracted from the target image to create a difference image. All non-varying sources are fully
subtracted in the difference image, leaving signal only for sources that have varied in brightness (or possibly position).

DIA is being increasingly used for precision photometry and transient detection in a wide range of photometric surveys
(e.g. PanSTARRS - \citealt{kai2002}, OGLE - \citealt{uda2003}, LSST - \citealt{ive2008}, RoboNet-II - \citealt{tsa2010}, QES - \citealt{als2013}, etc.).
Furthermore, these surveys are investing substantial efforts into post-calibration in order to minimise the systematic noise
in the survey photometry which affects important aspects such as the detection limits, homogeneity and completeness
(e.g. \citealt{stu2010}, \citealt{ofe2012}, \citealt{wit2012}, etc.).
Therefore, it is crucial to understand how the difference images, on which the photometry is performed, are created and
how systematic errors in the difference images themselves can affect the photometry. However, there is no study in the literature
on the systematic errors specific to DIA. In this research note, we have opened the investigation into systematic errors in DIA by exploring
the effect that an error in the kernel sum, known as the photometric scale factor, may have on the photometry.

\section{Equations}

The target image $I$ is modelled as the convolution of a reference image $R$ with
a convolution kernel plus a differential background. If one considers the kernel
as the product of a photometric scale factor $P$ and a normalised (by its sum) kernel $K$,
then the model target-image $M$ is defined as follows:
\begin{equation}
M = P \, (R \otimes K) + B
\label{eqn:model}
\end{equation}
The corresponding difference image $D$ is the image of model residuals given by:
\begin{equation}
\begin{aligned}
D & = I - M                      \\
  & = I - P \, (R \otimes K) - B \\
\end{aligned}
\label{eqn:diffim}
\end{equation}

Now assume that an object, consisting of a source of interest and a blend, has a true flux $f(t)$ on the photometric
scale of the reference image that is given by:
\begin{equation}
f(t) = \left( 1 + \frac{k(t)}{1 + k_{b}} \right) (1 + k_{b}) \, f_{S} \\
\label{eqn:ft}
\end{equation}
where $k(t)$ is a function of time $t$ that represents any variability in the source and $k_{b}$ is the true blend ratio.
Without loss of generality, adopt $k(0) = 0$ so that the quantity $f_{S}$ represents the true source flux at $t = 0$.
The true blend flux $k_{b} \, f_{S}$ makes the source appear brighter by a factor of
$(1 + k_{b})$ and reduces the apparent fractional flux-amplitude of any source variability by the same factor.

It follows that the true object flux $f_{R}$ on the reference image taken at $t = 0$ is:
\begin{equation}
f_{R} = (1 + k_{b}) \, f_{S}
\label{eqn:fr}
\end{equation}
and that the true object flux $f_{I} (t)$ on the target image taken at time $t$ is:
\begin{equation}
f_{I} (t) = P \, f(t)
\label{eqn:fi}
\end{equation}
By considering equation~\ref{eqn:diffim}, the true object flux $f_{D} (t)$ on the
difference image may be computed as:
\begin{equation}
f_{D} (t) = f_{I} (t) - P \, f_{R} \\
\label{eqn:fd}
\end{equation}
By substituting equation~\ref{eqn:fi} into equation~\ref{eqn:fd} and rearranging, $f(t)$ may be written
in terms of the measurable quantities $f_{R}$ and $f_{D} (t)$ as:
\begin{equation}
f(t) = f_{R} + f_{D} (t) / P
\label{eqn:ft2}
\end{equation} 
This is the equation used to convert difference fluxes to total fluxes and thence magnitudes
(e.g. \citealt{bra2011}).

Unfortunately, the modelling of the target image is never perfect and the difference image
consequently suffers from small systematic errors that are propagated to
the photometry. An error in the fitted differential background which results in a
non-zero background in the difference image is trivially accounted for, at the expense of a little
extra variance in the photometry, by including the background as a parameter in the method used
for performing the photometry on the difference image (i.e. PSF fitting or aperture photometry).
However, an error in the PSF matching (both shape and scale) produces systematic residuals at
the object positions that are more difficult to mitigate at the image processing stage.
Aperture photometry is agnostic to mismatches in PSF shape, but will be affected by an error
in the fitted photometric scale factor.
PSF photometry on the other hand is sensitive to mismatches in both PSF shape and scale
but will provide photometry with smaller variance than aperture photometry when stochastic
noise dominates. For the above reasons, we ignore the error in the fitted differential background,
we treat the error in the fitted normalised kernel as part of the error
introduced into the photometry by the measurement process on the difference image, and we assume
that the fitted photometric scale factor suffers from a small fractional error.

We use $P^{\,\prime}$, $K^{\,\prime}$, and $B^{\,\prime}$ to represent the fitted photometric scale factor,
kernel, and differential background, respectively. Employed in equation~\ref{eqn:diffim}, these give us the
difference image $D^{\,\prime}$ with systematic errors. If we assume that the difference flux is measured
with an error of $\varepsilon_{D} \, f_{I} (t)$ due to stochastic noise and/or the error in $K^{\,\prime}$,
and if we also assume that the error in $B^{\,\prime}$ can be successfully accounted for, then we obtain
the following {\it measured} difference flux on $D^{\,\prime}$ for the object:
\begin{equation}
f^{\,\prime}_{D^{\,\prime}} (t) = f_{I} (t) - P^{\,\prime} f_{R} + \varepsilon_{D} \, f_{I} (t)
\label{eqn:fdiff_meas}
\end{equation}
Adopting the expression $P^{\,\prime} = (1 + \varepsilon_{P}) \, P$ for the relation between $P^{\,\prime}$ and $P$,
and using equations~\ref{eqn:ft}~to~\ref{eqn:fi}, we may derive:
\begin{equation}
f^{\,\prime}_{D^{\,\prime}} (t) = P \, ((1 + \varepsilon_{D}) \, k(t) + (\varepsilon_{D} - \varepsilon_{P}) \, (1 + k_{b})) \, f_{S}
\label{eqn:fdiff_meas2}
\end{equation}
If we now assume that the method of performing photometry on the reference image yields a fractional
flux error of $\varepsilon_{R}$ in the reference flux, different to $\varepsilon_{D}$ because of the different nature of the
reference image and/or method used, then the measured reference flux is 
$f^{\,\prime}_{R} = (1 + \varepsilon_{R}) \, f_{R}$. Using $f^{\,\prime}_{R}$, $f^{\,\prime}_{D^{\,\prime}} (t)$
and $P^{\,\prime}$ in equation~\ref{eqn:ft2}, and doing some algebra, yields the following
expression for the measured object flux $f^{\,\prime} (t)$ on the photometric scale of the reference image:
\begin{equation}
f^{\,\prime} (t) = \left( 1 + \frac{k(t)}{(1 + \delta) \, (1 + k_{b})} \right) \, \left( \frac{1 + \varepsilon_{D}}{1 + \varepsilon_{P}} \right) \, (1 + \delta) \, (1 + k_{b}) \, f_{S}
\label{eqn:ft2_meas}
\end{equation}
where:
\begin{equation}
\delta = \varepsilon_{R} \left( \frac{1 + \varepsilon_{P}}{1 + \varepsilon_{D}} \right)
\label{eqn:delta}
\end{equation}
Equation~\ref{eqn:ft2_meas} describes how the measured flux of a constant source ($k(t) = 0$ for all $t$),
or a variable source ($k(t) \ne 0$ for at least some $t$), is distorted by the errors $\varepsilon_{D}$, $\varepsilon_{P}$ and $\varepsilon_{R}$.
The equation has been written in the form above to facilitate direct comparison to equation~\ref{eqn:ft} representing
the true object flux. The ratio of the measured to the true object flux is:
\begin{equation}
\frac{f^{\,\prime} (t)}{f(t)} = \left( 1 + \frac{\delta}{1 + \frac{k(t)}{1 + k_{b}}} \right) \, \left( \frac{1 + \varepsilon_{D}}{1 + \varepsilon_{P}} \right)
\label{eqn:ft_ratio}
\end{equation}

In magnitudes, equations~\ref{eqn:ft}~and~\ref{eqn:ft2_meas} become:
\begin{equation}
m(t) = -2.5 \log(f_{S}) - 2.5 \log \left( 1 + \frac{k(t)}{1 + k_{b}} \right) - 2.5 \log( 1 + k_{b} )
\label{eqn:mt_true}
\end{equation}
\begin{equation}
\begin{aligned}
m^{\,\prime} (t) = & - 2.5 \log(f_{S}) - 2.5 \log \left( 1 + \frac{k(t)}{(1 + \delta) \, (1 + k_{b})} \right)  \\
                   & - 2.5 \log(1 + \varepsilon_{D}) + 2.5 \log(1 + \varepsilon_{P})                           \\
                   & - 2.5 \log(1 + \delta) - 2.5 \log(1 + k_{b})                                              \\
\end{aligned}
\label{eqn:mt_meas}
\end{equation}
where $m(t)$ and $m^{\,\prime} (t)$ are the true and measured object magnitudes, respectively, on the magnitude scale of the reference image.

\section{Discussion}

The difference flux $f_{D} (t)$ is a quantity that is measured for each object on each difference image, and therefore $\varepsilon_{D}$
is specific to the object and difference image under consideration. However, for multiple difference images,
any systematic (as opposed to stochastic) component in $\varepsilon_{D}$ that is a function of either an object property (e.g. colour)
and/or an image property (e.g. pixel coordinates) may be estimated by solving for the appropriate magnitude offsets using the DIA photometry
of all of the constant objects in the corresponding target images. This is the approach, developed by authors such as \citet{hon1992}
and \citet{man1995}, that is starting to be adopted by many surveys as the standard procedure for performing a post-calibration of the
photometric data (e.g. \citealt{pad2008}). In this respect,
post-calibration of DIA photometry is no different than the post-calibration of photometry performed directly on the target images.
The appropriate magnitude offsets to be determined from the constant objects are represented by the term $- 2.5 \log(1 + \varepsilon_{D})$ in equation~\ref{eqn:mt_meas}
and their absolute values are usually of the order of $\sim~1~-~30$~mmag.

The reference flux $f_{R}$ is a quantity that is measured for each object on the reference image. Therefore the error
$\varepsilon_{R}$ is independent of the target image (or time) and it affects the photometry of constant objects
by making them systematically too bright ($\varepsilon_{R} > 0$) or too faint ($\varepsilon_{R} < 0$). Variable objects
suffer this same systematic error and in addition their fractional flux-amplitude of variation is either systematically amplified
($\varepsilon_{R} < 0$) or reduced ($\varepsilon_{R} > 0$) by a factor of $(1 + \delta) \sim (1 + \varepsilon_{R})$
to first order. In this respect, the effect of $\varepsilon_{R}$ is equivalent to that of an extra blend flux. The absolute value of
$\varepsilon_{R}$ is usually in the range of typical photometric precisions of $\sim~0.1~-~5$\%. Unless the DIA photometry has
a very small stochastic noise component ($\la$0.01\%), $\varepsilon_{R}$ is indistinguishable from $k_{b}$ since its effect on
$f^{\,\prime} (t)$ only differs from that of $k_{b}$ to second order (see equations~\ref{eqn:ft2_meas}~\&~\ref{eqn:delta}).
We do not consider any further the intricacies of disentangling the source flux, blend ratio and reference flux error.
Simply we note that when the blend ratio can be determined (e.g. by using external information, or for certain types of variability such as microlensing events),
the estimated blend ratio $k^{\,\prime}_{b}$ is related to the true blend ratio by $k^{\,\prime}_{b} \sim (1 + \varepsilon_{R}) \, k_{b} + \varepsilon_{R}$.

When fitting the model target-image, the photometric scale factor $P$ is typically assumed to be spatially-invariant and hence characterised by a single number,
although $P$ may also be modelled as a function of detector coordinates (\citealt{bra2013}). Regardless of how $P$ is modelled, the error $\varepsilon_{P}$
will be different for each target image since it is determined on a per-image basis from noisy images. Other effects such as flat-fielding errors, or changing non-uniform
atmopsheric extinction (i.e. clouds and/or airmass gradients),
may conspire to make $\varepsilon_{P}$ specific to the object and target image under consideration. As with $\varepsilon_{D}$, some of the systematic components
in $\varepsilon_{P}$ may be estimated by solving for the appropriate magnitude offsets using the DIA photometry of all of the constant objects in the
target images. In fact, the systematic components in $\varepsilon_{P}$ that are also common to those in $\varepsilon_{D}$ may be absorbed into the
magnitude offsets represented by the term $- 2.5 \log(1 + \varepsilon_{D})$ in equation~\ref{eqn:mt_meas}. However, the remaining systematic components
in $\varepsilon_{P}$, if left uncorrected, will cause errors in the DIA photometry that have the potential to be misinterpreted as real signals.

\section{Summary And Recommendations}

The vast majority of photometric surveys employing DIA do not consider $P$ to be a source of error since it is
assumed that $P$ has been precisely determined, which is equivalent to assuming that the differences in atmospheric extinction between the
reference and target images have been perfectly accounted for by the DIA modelling itself. However, in this research note we have shown 
that any errors in $P$ that do exist will have an important impact on the DIA photometry. We therefore strongly recommend that it
becomes standard procedure to assess, and if necessary correct for, the effect of the mean error in $P$ for each target image
on the DIA photometry (encapsulated by the term $2.5 \log(1 + \varepsilon_{P}(t))$ in equation~\ref{eqn:mt_meas}).

One method to do this involves fitting a post-calibration photometric model including a set of per-image magnitude offsets
$2.5 \log(1 + \varepsilon_{P}(t))$ to the DIA photometry of all of the constant objects in the target images.
Note that the post-calibration model should also include the magnitudes of the constant objects as free parameters
(see \citealt{bra2012}) and any other relevant terms such as $- 2.5 \log(1 + \varepsilon_{D})$.
If the variations in the per-image magnitude offsets are found to be smaller than the level of the stochastic noise
in the best object photometry, then they may be dropped from the post-calibration model. However, if they are 
deemed to be significant, then the per-image magnitude offsets $2.5 \log(1 + \varepsilon_{P}(t))$ may
be used to correct the DIA photometry of all of the objects for the photometric error introduced by $\varepsilon_{P}(t)$.

This technique has started to be used in the series of papers on variable stars in globular clusters by the
lead author (e.g. \citealt{kai2013}, \citealt{are2013}, etc. which employ the methodology of \citet{bra2012}), and in these works
the relevant per-image magnitude offsets are found to be at the 0.1-2\% level (see Figure~1 of \citealt{kai2015}).
This example clearly demonstrates that DIA does not always perform the photometric matching between images to a precision that is below the
stochastic noise in the photometric measurements and it serves to emphasise how crucial it is to account for the mean error in $P$
for each target image in order to minimise the associated systematic errors.

Finally we caution that when low-level ($\la$1-2\%) suspected signals occur in the DIA photometry of an object that cannot be confirmed either
by the detection of a repeating signal (for periodic signals) or by independent observations taken at the same epoch, a careful analysis of the images
on which the suspected signal was detected is warranted. Checks should be performed for the presence of clouds (e.g. light cirrus) that may
have caused non-uniform atmospheric extinction across the field-of-view that also evolves throughout the time-series with the cloud movement
since this can cause smooth temporal variations in $\varepsilon_{P}$ that are different for each object and that will manifest themselves as
smooth variations in the object light curves. Adopting a spatially variable photometric scale factor in the model target-image for the DIA may
partly mitigate this problem.

\begin{acknowledgements}
This publication was made possible by NPRP grant \# X-019-1-006 from the Qatar National Research Fund (a member of Qatar Foundation).
The statements made herein are solely the responsibility of the authors.
\end{acknowledgements}

%-------------------------------------------------------------------

\end{document}